# Implicit Spatiotemporal Bandwidth Enhancement Filter by Sine-activated Deep Learning Model for Fast 3D Photoacoustic Tomography

I Gede Eka Sulistyawan, Takuro Ishii, Riku Suzuki, and Yoshifumi Saijo

*Abstract*—3D photoacoustic tomography (3D-PAT) using high-frequency hemispherical transducers offers near-omnidirectional reception and enhanced sensitivity to the finer structural details encoded in the high-frequency components of the broadband photoacoustic (PA) signal. However, practical constraints such as limited number of channels with bandlimited sampling rate often result in sparse and bandlimited sensors that degrade image quality. To address this, we revisit the 2D deep learning (DL) approach applied directly to sensor-wise PA radio-frequency (PARF) data. Specifically, we introduce sine activation into the DL model to restore the broadband nature of PARF signals given the observed band-limited and high-frequency PARF data. Given the scarcity of 3D training data, we employ simplified training strategies by simulating random spherical absorbers. This combination of sine-activated model and randomized training is designed to emphasize bandwidth learning over dataset memorization. Our model was evaluated on a leaf skeleton phantom, a micro-CT-verified 3D spiral phantom and *in-vivo* human palm vasculature. The results showed that the proposed training mechanism on sine-activated model was well-generalized across the different tests by effectively increasing the sensor density and recovering the spatiotemporal bandwidth. Qualitatively, the sine-activated model uniquely enhanced high-frequency content that produces clearer vascular structure with fewer artefacts. Quantitatively, the sine-activated model exhibits full bandwidth at -12 dB spectrum and significantly higher contrast-to-noise ratio with minimal loss of structural similarity index. Lastly, we optimized our approach to enable fast enhanced 3D-PAT at 2 volumes-per-second for better practical imaging of a free-moving targets.

*Index Terms*—Photoacoustic imaging, Deep learning, Radio frequency

## I. INTRODUCTION

Photoacoustic tomography (PAT) has increasingly become a popular modality for spectroscopic quantification [1], elucidating a variety of metabolism such as oxygenation. Aiming to probe such physiological dynamics at finer and more complex anatomical structure, researchers have advanced high-frequency PAT systems from 2D rotational scanning of one-dimensional array transducers to real-time 3D volumetric imaging with two-dimensional matrix array transducers [2], [3], [4], [5]. Although, our current design of two-dimensional matrix array transducers faces challenges in addressing the broadband and omnidirectional wave propagation nature of photoacoustic waves [6]. Firstly, high-frequency reception of the broadband photoacoustic signal is critical as they carry the shorter wavelength for finer image resolution. However, as of now, these high-frequency PAT systems remain bandlimited, typically up to 20 MHz bandwidth [7], [8], [9], [10]. Secondly, omnidirectional waves like the photoacoustic ideally require a large and spherical aperture, but the impracticality of enclosed spherical transducers and limited channels on the acquisition hardware often led to sparse sensor placement on hemispherical arrangement. Shown in Fig. 1, the combination of sparse and band-limited sensors results in enlarged and elongated target representations. These limitations highlight the need for advanced processing techniques to effectively compensate for the constraints of current PAT systems.

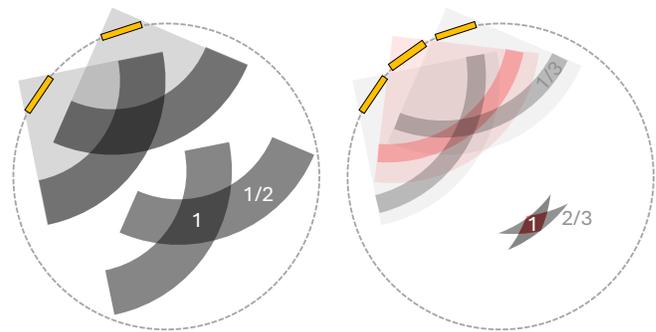

Fig. 1. Impact of sparse bandlimited sensors on imaging (left) and its enhanced version (right) with one more sensor in between at higher bandwidth. The number is the relative intensity with threshold at 1/3.

As reviewed in [11], deep learning (DL) has emerged as a popular alternative to conventional techniques for improving visualization quality in PAT systems. However, the existing models are primarily designed for envelope-intensity-based image-to-image enhancement, which introduces two significant challenges: (1) limited scalability to 3D volume-to-volume enhancement due to computational cost of 3D operations (e.g., convolutions) and (2) the requirement for large training datasets, as DL models must learn to generalize across a wide

This work is supported by JST-BOOST, Japan Grant Number JPMJBS2421.

I.G.E. Sulistyawan, T. Ishii, R. Suzuki and Y. Saijo are with the Graduate School of Biomedical Engineering, Tohoku University, Sendai 980-8579, Japan. (Corresponding author: Y. Saijo email: saijo@tohoku.ac.jp).

variety of anatomical structures. Several studies have attempted direct volume-to-volume enhancement [12], [13]. However, the processing was done offline, highlighting its high computational demands in 3D operations. The computational burden of volume-to-volume enhancement could be alleviated by adopting a simpler 2D enhancement approach on sensor-wise photoacoustic radio frequency signals (PARF) since it offers a compact 2D representation of the 3D volume.

The sensor-wise PARF enhancement has been attempted in the following three studies. A work by Awasthi et al., utilized DL to address the limited sensor count and bandwidth of the sensor-wise PARF via UNET architecture with hybrid rectified linear unit (ReLU) and exponential linear unit (ELU) [14]. Sharing similar objectives, Yamakawa et al., extended the DL technique to sensor-wise PARF enhancement for 3D PAT [15]. A custom ResNet was trained with ReLU activation function and capable of interpolating radio frequency signals between two sensors to reduce the artefact. However, the DL did not address the bandwidth-limit problem. Finally, the latest work by Li et al. proposed a diffusion model to address the limited view of the PAT [16]. However, the diffusion model took additional iterations of ~ 6 minutes to complete and did not consider bandwidth-limit problem. Therefore, there is still a wide room for improvement to enable efficient sensor-wise PARF enhancements on 3D-PAT, particularly in terms of real-time realizability, addressing both limited-bandwidth and sparse sensor in one DL.

Another challenge in volume-to-volume enhancement is the need for large training datasets to capture diverse structures. The k-Wave simulation [17] is commonly used to generate training PARFs dataset but 3D simulations are computationally expensive, and reference datasets remain limited [18], [19], [20]. To alleviate the burden in simulating PARF, we found the potential of a former study by Guo et al that suggested the PARF could be represented as superpositions of some spherical absorbers [21] which later degraded by system constraints such as the sensor's bandwidth [9]. In line with this idea, further exploration is needed to evaluate whether this simpler model can effectively replace k-Wave for generating training data.

Aiming to advance our current state of DL to alleviate the sparsity of the sensors and limited bandwidth problems, in this paper, we propose the following frameworks:

1. A PARF simulation framework for a 3D PAT system using a random spherical absorber model as a structure-independent training data generator that let DL training solely focuses on PARF enhancement.

2. A spatiotemporal bandwidth enhancement DL framework using sine activation function with the hypothesis that such a function may enable the DL model to better learn the features of a sensor-wise PARF, particularly since the raw radio frequency data on the sensor-wise PARF resembles raw photoacoustic data that are rich in broadband features.

## II. MATERIAL & METHODS

This section consists of three parts. First, part A provides the specification of our 3D-PAT system and a simulation setup to generate structure-independent PARF dataset of the 3D-PAT system. Second, in part B, we introduce our DL framework using a sine activation function and a training strategy using the dataset generated in part A. Third and finally, part C outlines our evaluation strategy for the developed DL model.

### A. Setup for Sparse Hemispherical 3D Photoacoustic Tomography (3D-PAT) and Its PARF Simulation Framework

#### A.1 The Hemispherical 3D-PAT System

Our group has established a high-center-frequency 3D photoacoustic tomography (3D-PAT) system [3] consisted of a custom-made hemispherical transducer (JAR811, Japan Probe Co., Ltd., Kanagawa, Japan) shown in Fig. 2A. The transducer comprised of semi-regularly-arranged 256 element sensors at seven concentric rings [23], and the sensors were placed on quarter-$\pi$ hemispherical geometry that had a focal point at 30 mm away from the surface of the sensors. The imaging field of view (FOV) at the focal point was 2 mm$^3$ presented with 86 x 86 x 86 voxels to satisfy the spatial Nyquist limit. Each sensor had a dimension of 2.2 mm $\times$ 2.2 mm, and the element pitch and kerf were ~2.44 mm and ~120 μm, respectively. The center frequency of each sensor was 12 MHz (Bandwidth 8 MHz to 16 MHz). The 3D-PAT was equipped with a 20 Hz pulse-repetition-rate optical parametric oscillator laser at visible and near infrared wavelengths (OPOTEK Opolette HE 355 LD, California, USA). In this study, we tuned the wavelength at 546 nm to visualize both artery and vein simultaneously and, thus, considering the most complex structure scenario *in vivo* [24]. The custom system was controlled by a research-purpose ultrasound platform (Vantage 256 High-frequency configuration, Verasonics Inc., WA, USA).

On each laser pulse, each sensor captures a PARF consisting of 256 time points sampled at 62.5 MS/s. Therefore, a single acquisition generates a data of 256-time samples for each of the 256 sensors. This PARFs arranged column-by-column is referred to as the sensor-wise PARF and is visualized as a 2D matrix, where the x-axis represents the sensor index, and the y-axis corresponds to the time samples. Finally, we performed laser-artefact filtering prior to any processing which we described detailly in Supp. Mat. A.

Considering that the wavelength of a 12 MHz acoustic wave in water (speed of sound is 1475 m/s) is approximately 122 μm, the kerf of our transducer (~120 μm) is large enough to lag approximately one phase of signal between adjacent PARF. Transducers with such wide spacing are often referred to as sparse sensor arrays [25] which are prone to streaking artefact. Therefore, interpolating an additional virtual sensor between each pair of existing sensors might effectively alleviate the sparsity-induced streaking artefact by reducing the kerf by half.

#### A.2 PARF Simulation with Sub-resolution Spherical Absorbers

One of the keys of this study is to simplify the training data generation approach. To do so, we designed a simulation framework inspired by Guo et al by assuming randomized point spherical absorbers in the imaging FOV [21]. Eq. (1) described $Q$ spherical absorbers with radius $R_q$ at coordinate $\vec{r}_q$ producing ideal photoacoustic wave $s_q$ which arrived at a sensor $k$ and were super-positioned into the ideal sensor-wise PARF, that is,

$$s_k^{ideal}(t) = \sum_{q \in Q} s_q(t; R_q, \vec{r}_q). \quad (1)$$

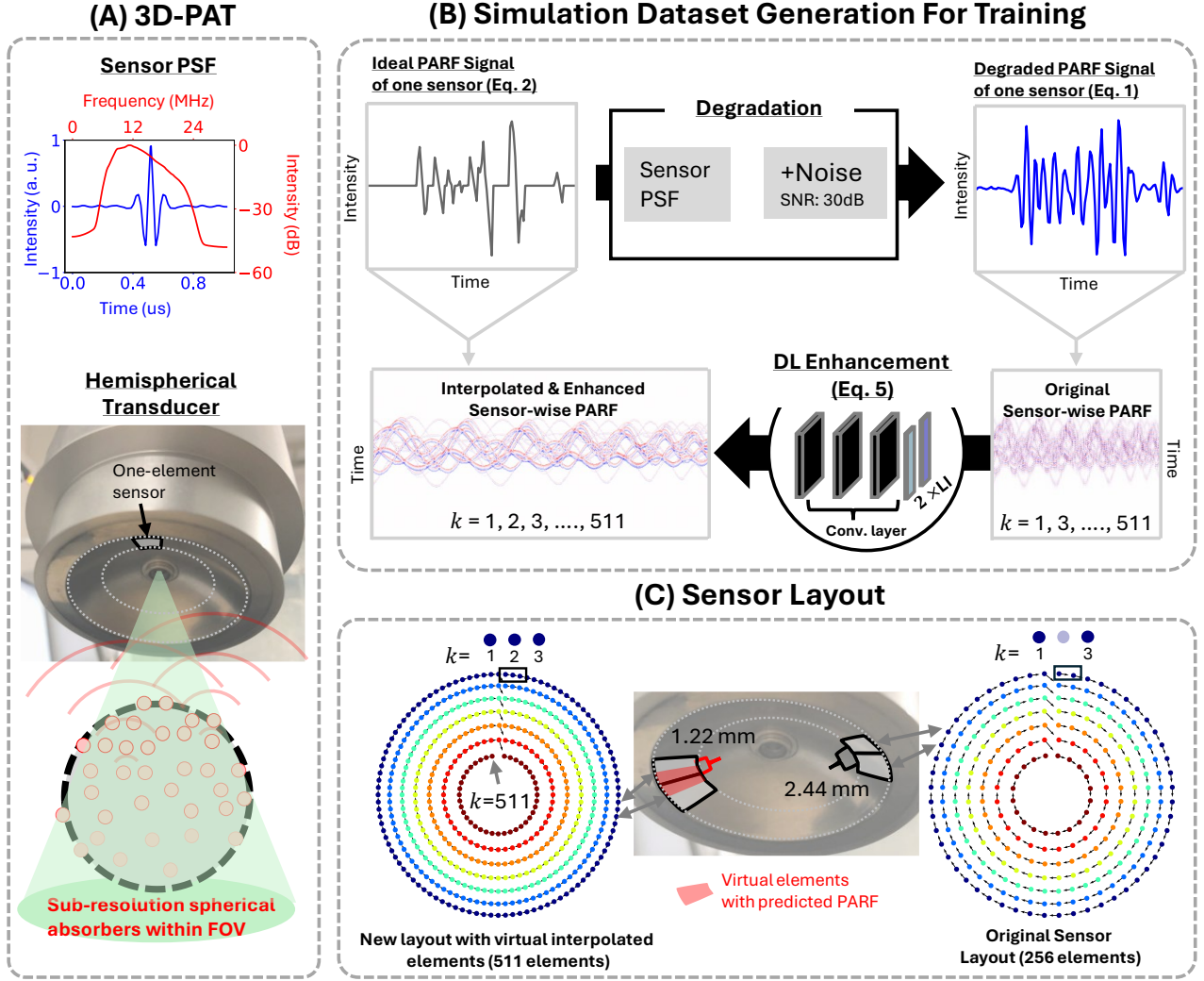

**Fig. 2.** (A) The properties of the 3D-PAT comprised of hemispherical transducer with the sensor PSF centered at 12 MHz. (B) Procedure to obtain simulation dataset for training from ideal signal generation, degrading procedure and the DL enhancement. LI is linear interpolation. (C) Relation between the indexing and their physical location on the transducer.

Here, $s_q$ is calculated from the ideal N-shape model [26] as follows,

$$s_q(t; R_q, \vec{r}_q) = \frac{r+ct}{2r} p_0(-r+ct) + \frac{r-ct}{2r} p_0(r-ct), \quad (2)$$

with,

$$r = \|\vec{r_k} - \vec{r_q}\| \quad (2a)$$

$$p_0(r) = U(r)U(-r + R_q) \quad (2b)$$

where, $c$ is speed of sound, $r$ is the distance between sensor $k$ and the spherical absorber $q$, and $U$ is unit step function.

The simulated observed PARF signal at sensor $k$ named $s_k^{obs}(t)$ is obtained by degrading the ideal sensor-wise PARF $s_k^{ideal}(t)$ by convolution with the sensor-specific point spread function (PSF) and added noise. The process is expressed by,

$$s_k^{obs}(t) = h(t) \star s_k^{ideal}(t) + n(t) \quad (3)$$

where $n(t)$ is white noise to make signal-to-noise ratio of 30 dB and $h(t)$ is the time domain of the PSF shown in Fig. 2A that is conformable with our previously report [3], [23].

The variable $k$ in the simulation referred to the index of the transducer with 511 sensors, where the odd indexes $k \in [1, 3, \ldots]$ are assigned to the physical elements in the original sparse transducer with 256 sensors. As shown in Fig. 2B, the simulation produced all the necessary data for training, i.e., the ground-truth data of the ideal sensor-wise PARF $s_k^{ideal}(t)$ and the input data of 256 sensor-wise PARF (thereafter called $s_{odd\,k}^{obs}(t)$) that can be practically acquired with the 3D-PAT system, respectively. Fig. 2C illustrates the relationship between the original and the twice-interpolated sensors, including their physical location on the actual transducers.

### B. Deep learning modeling and training

The goal of DL is to obtain $s_k^{pred}(t)$ that approximates the ideal $s_k^{ideal}(t)$ given the observed sensor-wise PARF at only the odd index, $s_{odd\,k}^{obs}(t)$. Part B.1 begins with the baseline method using the conventional time-domain Wiener filter. Then, we described the explored DL architectures and the training strategies in part B.2 Finally, we detailed the use of sine activation function within the DL models in part B.3.

*B.1 Conventional Time-domain Wiener Filter*

We have reported the use of Wiener filter to accommodate the limited bandwidth of the sensor [27]. In this study, we employed Wiener filter as a baseline to enhance the bandwidth of the original sensor which briefly described as,

$$s_{\text{odd }k}^{pred}(t) = s_{\text{odd }k}^{obs}(t) \star m(t) \qquad (4)$$

where,

$$m(t) = \mathcal{F}^{-1}\left( \frac{H^*(f)}{|H(f)|^2 + \beta \frac{N(f)}{S(f)}} \right) \qquad (4a)$$

with, $\mathcal{F}^{-1}$ is inverse 1D time-domain Fourier transform, $H$ is the frequency spectrum of the PSF obtained in (3), $s_{\text{odd }k}^{obs}(t)$ is the acquired PARF signal at the original sensor, $N(f)/S(f)$ is signal to noise ratio, and $\beta$ is hyperparameter. Throughout this study, signal to noise ratio is 30 dB and $\beta$ is fixed at 0.01. Note that conventional Wiener filter did not interpolate the sensor and thus the even index, $s_{\text{even }k}^{pred}(t)$, is simply zero.

*B.2 Deep Learning Models and Training*

Our DL model replaces Eq. (4) with a parameterized spatiotemporal enhancement filter that is built upon a fundamental building block described as:

$$\mathbf{S}^{pred} = f(\mathbf{W} \star \mathbf{S}^{LI} + b), \qquad (5)$$

where, $\mathbf{W}$ is trainable 2D filter, $\star$ is 2D convolution operator, $b$ is bias, $f$ is activation function, and $\mathbf{S}^{LI}$ the output of sensor-wise linear interpolation applied to $\mathbf{S}^{obs}$. Instead of time-domain operation as previously used in (4), the matrix notation on (5) indicates that the PARF is treated as a spatial-temporal data (see Fig. 2B Sensor-wise PARF). Particularly, $\mathbf{S}^{pred}$, $\mathbf{S}^{obs}$ and $\mathbf{S}^{ideal}$ are the matrix form of $s_k^{pred}(t)$, $s_k^{obs}(t)$ and $s_k^{ideal}(t)$ respectively.

Using Eq. (5), we constructed three architectures: UNET [14], Fully dense (FD) UNET [28] and ResNet [15]. From [14], we adopted the UNET architectures and the Relu/ELU activation function with linear sensor-wise interpolation without PARF patching. The rest models, i.e., FD-UNET [28] and ResNet [15], were implemented following the previous studies. Table I presented a brief overview of all models investigated with the detailed architecture presented in Supp. Mat. B. Lastly, the UNET serves as our primary architecture which will be modified with a variety of activation functions to be described in B.3.

For training, PARF of four random spheres was simulated for each iteration with positive radius $R_q$ drawn from a folded normal distribution of $50 \pm 45$ μm, approximated upon the strongest signal emitted at radius ~41 μm according to $0.33c/f_{peak}$ (Fig. 1 in [29]) and the maximum wavelength of our transducer at 92 μm. The coordinate $\vec{r}_q$ was drawn from uniform distribution within the FOV. The training objective was mean squared error (MSE) between $\mathbf{S}^{pred}$ and $\mathbf{S}^{ideal}$. We trained all models using Adam optimizer for 100,000 iterations with 5E-6 learning rate and 200,000 more fine-tuning iterations with 1E-6 learning rate. The DL development and training was conducted on PyTorch 2.8 running on a workstation equipped with a NVIDIA RTX 5080 (16 GB memory, CUDA 12.8).

Table I. Brief summary of the architectures.

| Model Name | Ref | Act. Func. | Input Range | Param. (MB) |
|---|---|---|---|---|
| UNET-SINE | This study | Sine | [-1, 1] | 118.37 |
| FD-UNET-SINE | | Sine | [-1, 1] | 58.27 |
| RESNET-RELU | [10] | ReLU | [0, 1] | 1.41 |
| FD-UNET-RELU | [24] | ReLU | [0, 1] | 58.27 |
| UNET-RELU-ELU | [9] | ReLU and ELU | [-1, 1] | 118.37 |
| UNET-MIRROR-RELU | [26] | Mirrored ReLU | [-1, 1] | 257.97 |

*B.3 Incorporating Sine as Activation Function*

Awasthi et al. reported that a combination of ReLU and ELU as activation functions is better at capturing negative intensity of the PARF [14]. On another related study, Lu et al. proposed the use of mirrored ReLU by concatenating the negative ReLU-activated intensity in addition to the positive ones. Such concatenation was expected to preserve both positive and negative information of the signal [30]. In this study, both ReLU/ELU and the mirrored ReLU are investigated under the same UNET model. Note that mentioned in [30], the use of concatenation was doubled the trained features and considered as an advantage, thus we kept it double.

To supplement the current knowledge, this study further explored the use of sine activation function as it may have better handling on high-frequency context [31]. We hypothesized that this property would offer a more suitable representation for the sensor-wise PARF. A simplified conceptual explanation of the potential impact of using sine as an activation function and its difference from other functions are illustrated in Fig. 3. The illustration provides useful intuition, suggesting that the sine function may "unwrap" the intensity as weights increase and thus, emphasizing higher-frequency features. However, it should be noted that this simplification may not fully reflect the true behavior as the relationship between the learned filters and the input is often highly non-linear [32].

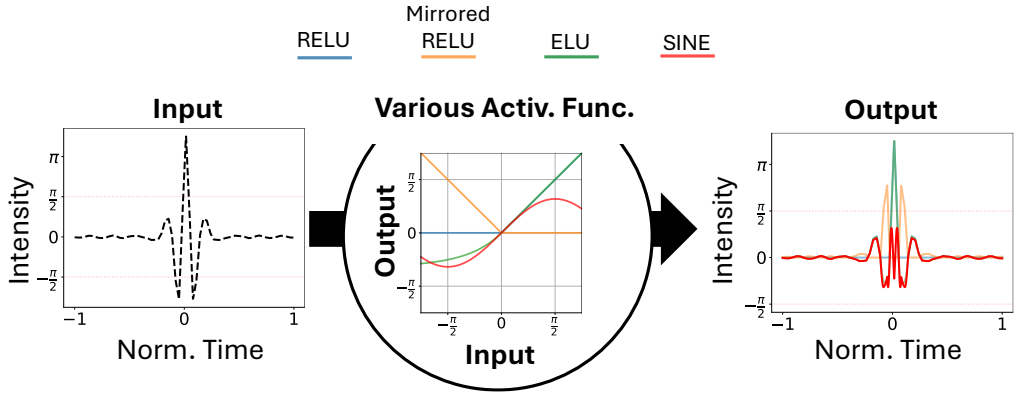

**Fig. 3.** Comparison of different activation functions.

## C. Signal-based and 3D Image-based Evaluation of the DL models

Prior to evaluation, we conducted a preliminary investigation to ensure that all models were properly trained despite the randomization. We also verified that the model utilized the periodicity of the sine activation function by confirming that the pre-activated features were distributed beyond the interval $\left[-\frac{\pi}{2}, \frac{\pi}{2}\right]$. Details of this preliminary evaluation are provided in Supp. Mat. C. After confirming proper model training, we proceed in C.1 to analyze the behavior of the trained model. Then, we describe the procedure for volume reconstruction in C.2 used for practical evaluation. Applied on the volume, we introduce the evaluation material and metrics in C.3.

### C.1. Analysis of DL Model's Behavior

To elucidate the generalized behavior of the model, this study employed a technique for probing the inductive bias of a DL model [33]. Specifically, Gaussian noise of the same dimensionality as $\boldsymbol{S^{obs}}$, normalized to the range of [-1, 1] was fed into the trained model. Any structure that emerged in the output was interpreted as the model's inductive bias. To further characterize these behaviors, the magnitude spectrum of the 2D-FFT of the output was analyzed on a logarithmic (dB) scale. Recall that the input to the model was the 2D sensor-wise PARF data where the x-axis represents spatial sensor position and the y-axis represents time. Therefore, the resulting 2D-FFT maps the spatial frequency in wavenumber along the x-axis and the temporal frequency along the y-axis. The contour of the magnitude spectrum at specific threshold, therefore, highlights the region of the highest spectral energy the model inherently reconstructs from pure noise.

### C.2. Volumetric Image Reconstruction

We used the UBP algorithm [34] to obtain the volume from the sensor-wise PARF. The UBP is defined as,

$$p(\vec{r}) = \sum_{k=1}^{N} s_k^{pred}(t - \Delta t_k) \quad (7)$$

with $s_k$ is the enhanced PARF at sensor $k$, $\Delta t_k$ is the time delay between the reconstructed coordinate $\vec{r}$ and the location of sensor $k$ at $\overline{r_k}$, calculated as follows,

$$\Delta t_k = \frac{\|\overline{r_k} - \vec{r}\|_2}{c}. \quad (7a)$$

The wave propagation was considered straight without refraction between the object and the coupling interface.

In addition to UBP, we employed the coherence factor (CF) weighing to avoid any noises other than the sparse sensor and limited bandwidth artefact that potentially biasing the evaluation [35]. We used an intensity-based CF defined as follows,

$$CF(\vec{r}) = \frac{1}{N} \frac{\left|\sum_{k=1}^{N} s_k^{pred}(t - \Delta t_k)\right|^2}{\sum_{k=1}^{N} \left|s_k^{pred}(t - \Delta t_k)\right|^2}, \quad (8)$$

with the normalizing denominator $N$ is the effective sensor number i.e., 256 for the conventional and 511 for the DL-enhanced. The final CF-weighed volume for practical evaluation then became,

$$p_{CF}(\vec{r}) = p(\vec{r}) \, CF(\vec{r}). \quad (9)$$

### C.3. 3D Image-based Evaluation

The volumetric images constreud in C.2 were utilized to further evaluate the practical performance of each DL model using multimodal reference phantoms. First, DL models are evaluated based on 2D maximum intensity projection (MIP) of a leaf phantom embedded in 7.5% transparent agar phantom. Second, we obtained a 3D reference by making a spiral phantom where the absorbance was made of a 10% barium sulfate suspension (Barium Sulfate, ReagentPlus 99%. Sigma Aldrich, Darmstadt, Germany) in black-inked 5% polyvinyl alcohol. The aqueous absorbance then injected into a spiral mold made inside a transparent 15% polyacrylamide phantom [36] and let froze for ~20 minutes then thawed. The composite of barium sulphate and black ink enabled comparing the 3D-PAT volume with Micro-CT [37]. In total, we obtained 1,595 leaf phantom patches to evaluate the 2D MIP and 870 spiral phantom volume patches to evaluate the 3D structure. For quantitative evaluations, we use the following metrics: structural similarity index (SSIM) and contrast to noise ratio (CNR). The contrast to noise ratio is defined as

$$CNR = \frac{\mu_{mask\,p_{CF}} - \mu_{outside\,mask\,p_{CF}}}{\sigma_{outside\,mask\,p_{CF}}} \quad (10)$$

where, mask is binarized region obtained from the reference that indicates the region where absorber existed. Notation $\mu$ and $\sigma$ are mean and standard deviation, respectively. The given CNR is simply the ratio of intensity between the inside of the known location of the absorber by the reference against the background intensity outside the mask. When reference is unapplicable, region of interest (ROI) will be specified. Lastly, SSIM is used to supplement the CNR in terms of potentially attenuating intensity and trading off fine structures.

Finally, *in vivo* qualitative evaluation on human superficial palm microvasculature was conducted. We optimized the trained model along with the rest of the signal processing into executable software for direct near-real-time testing on the 3D-PAT system. The signal processing was done on a workstation (equipped with RTX Ada 2000) that was directly connected to the Vantage ultrasound platform. In detail, the online signal processing was conducted every 10th laser pulse (500 ms). Therefore, the effective volume rate was 2 Hz accounting all processes from DL enhancement to visualization. Among these timelapse volumes, one will be selected for models' evaluation.

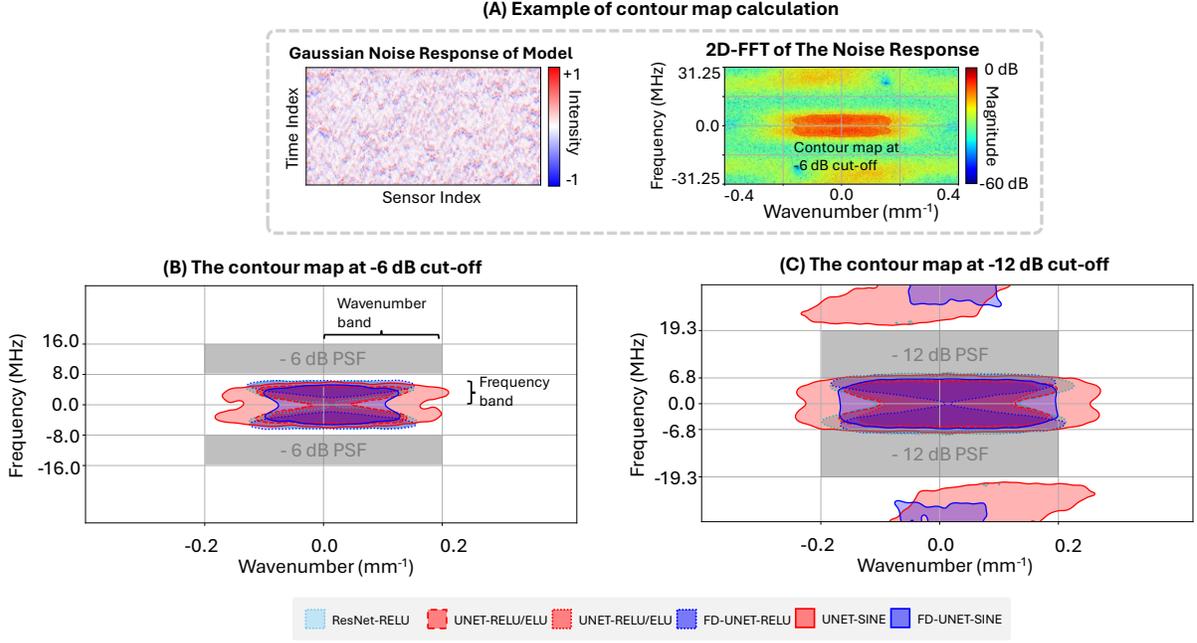

**Fig. 4.** Inductive bias elicited by each model presented as -6 dB and -12 dB bandwidth. (A) Output UNET-SINE given input Gaussian noise input and example of obtaining the spatiotemporal spectrum by 2D-FFT. An example of -6 dB contour drawing is presented on the 2D-FFT. (B) and (C) Compares the contour at -6 dB and -12 dB cut-off, respectively. Gray area is the theorized cut-off of the ideal PSF of the transducer.

### III. RESULTS

#### A. Model Characteristic

Fig. 4A displayed an example of the UNET-SINE response pattern given Gaussian noise input. On Fig. 4A, we showed an example of obtaining the spatiotemporal spectrum using 2D FFT of the DL's output given Gaussian noise. Later, we analyze the contour map of the spatiotemporal spectrum at cut-off -6 dB and -12 dB. Shown on Fig. 4B, we observed that the trained DL had strong biases towards amplifying the low-frequency spectrum at up to 8.0 MHz of the temporal bandwidth and 0.2 mm⁻¹ wavenumber bandwidth. Lowering the cut-off to -12 dB shown on Fig. 4C, we started to observe the implicit behavior of the sine-activated DL to also amplify the high-frequency spectrum. These two observations became qualitative empirical evidence where the DLs had attempted to broaden the spatiotemporal bandwidth.

Based on the -6 dB cut-off on Fig. 4B, the ResNet-ReLU showed the widest low-frequency amplification up to 6.5 MHz and followed by UNET-SINE at 6 MHz. At the -12 dB cut-off on Fig. 4C, both models reached similar low-frequency enhancements up to 8.3 MHz. However, only sine-activated model enhanced the high-frequency spectrum, particularly between 20 – 31.25 MHz obtained by the UNET-SINE model. This suggested that the sine-activated model could better capture the fine details of the PARF. However, all models produced similar wavenumber bandwidth except for UNET-SINE which slightly extended to 0.25 mm⁻¹. This minor improvement was likely due to the limited sensor layout at quarter-π arrangement. Overall, UNET-SINE provided the best performance by more effectively extending both temporal frequency bandwidth and spatial wavenumber bandwidth beyond the limit of the original transducer.

#### B. Phantom Evaluation

##### B.1 2D Leaf skeleton phantom evaluation

Fig. 5 compares DL reconstructions of a leaf skeleton phantom using reference optical image. All models effectively suppress streaking artifacts but tend to miss fine details, e.g., the ~70 μm skeleton in region 1 on Fig. 5 being considered in the training distribution. Elucidating more into the missing small skeletons, Fig. 6 showed that UNET-SINE captures the small skeletons before weighing CF, but these skeletons disappeared afterward which suggests CF suppresses both noise and the skeletons. Structures near the vicinity of the FOV also vanished on Fig. 6 after linear interpolation and was found partially recovered by UNET-SINE.

Fig. 7 displayed the graph summary of quantitative evaluation of the leaf phantom skeleton experiment. In general, we observed that using CF greatly improves the imaging quality in terms of artefact reduction and conformability with reference. Thus, we made comparisons with the highest performance of each method. The UNET-SINE had a significantly higher CNR at 0.261 (± 0.34) against the secondly performed FD-UNET-SINE with CNR of 0.209 (± 0.31) which is insignificantly higher than the third one achieved by UNET-MIRROR-RELU with CNR of 0.204 ( ± 0.32). This significance was likely achieved by the UNET-SINE as it suppresses the leaking intensity near the bifurcating area shown on region 2 Fig. 5. The FD-UNET-SINE and UNET-SINE yielded the second and third best SSIM at 0.356 (± 0.22) and 0.356 ( ± 0.22), respectively. Although, these SSIMs had insignificant differences against the best SSIM obtained by UNET-MIRROR-RELU at 0.360 (± 0.22). Thus, we confirmed that the sine-activated model had performed better than its non-sine-activated competitors at artefact reduction with insignificant trade-off in structure.

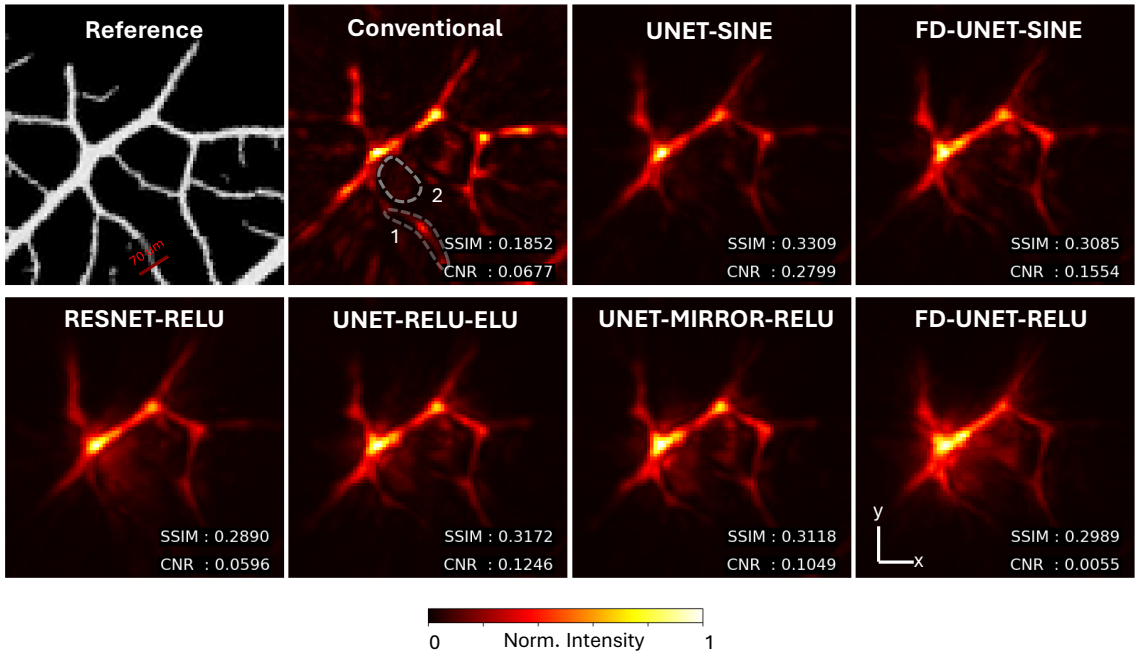

**Fig. 5.** Leaf phantom experiment across different model. Reference is the binarized optical image of the leaf skeleton. Region 1 highlighted missing small-diameter skeleton. Region 2 highlighted leaking intensity around bifurcating area. XY scalebar is 250 um.

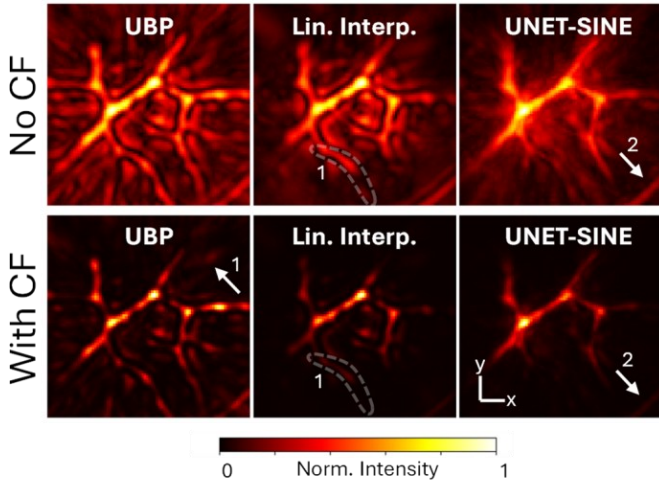

**Fig. 6.** Comparison before and after CF. Region 1 is the same with Fig. 5. Arrow 1 pointed at remained streaking artefact, Arrow 2 pointed at missing structure at the vicinity of FOV. Scalebar 250 um.

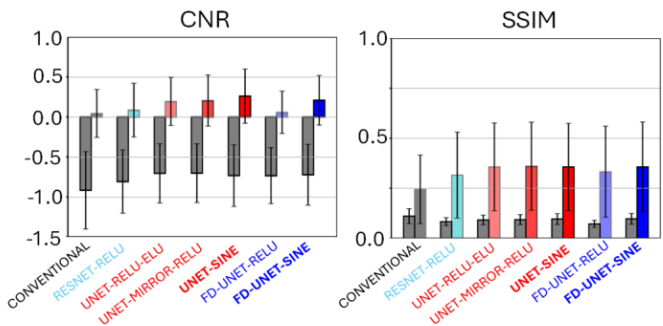

**Fig. 7.** Quantitative evaluation of leaf phantom study. Gray and colored graph are without and with CF, respectively.

### B.2 3D Spiral phantom evaluation

Moving forward to 3D visualization, Fig. 8 displays the rendered volumetric view of the spiral along with the Micro-CT reference and its optical picture. In general, we observed that our 3D-PAT system with quarter-$\pi$ hemispherical viewing angle missing the segment of the spiral perpendicular to the transducer's aperture i.e., the z-axis vertical component. This can be clearly visible by taking the XZ-plane MIP where most of the reconstructed image missed the vertical structure. In terms of artefact reduction, the UNET with sine activation function performs slightly better than others by attenuating the artefact visible on the pointed arrow and focusing the intensity on the spiraling structure.

Fig. 9 displayed the graph summary of the quantitative evaluation of the spiral phantom experiment. For this case, the negative CNR was still inevitable due to the missing vertical component as consequence of the quarter-$\pi$ hemispherical viewing angle. In general, both sine-activated models yielded the best two performing models. Particularly, UNET-SINE being the best performing in both CNR and SSIM with value of -0.069 ($\pm$ 0.01) and 0.955 ($\pm$ 0.01). In accordance with the image evaluation, the volume evaluation also showed the UNET-SINE to be a worthy model to enhance the PARF.

### C. In-vivo Evaluation

Upon successful result in the phantom study, we moved forward to an *in vivo* evaluation with online DL processing. Video on supplementary material 1 was taken while doing *in vivo* imaging at 2 volume-per-second with the enhanced volume reconstruction, proving that our DL approach was executable for fast 3D visualization. As shown, we were able visualize the microvasculature as the palm was moving and thus, potentially able to probe a specific location with ease.

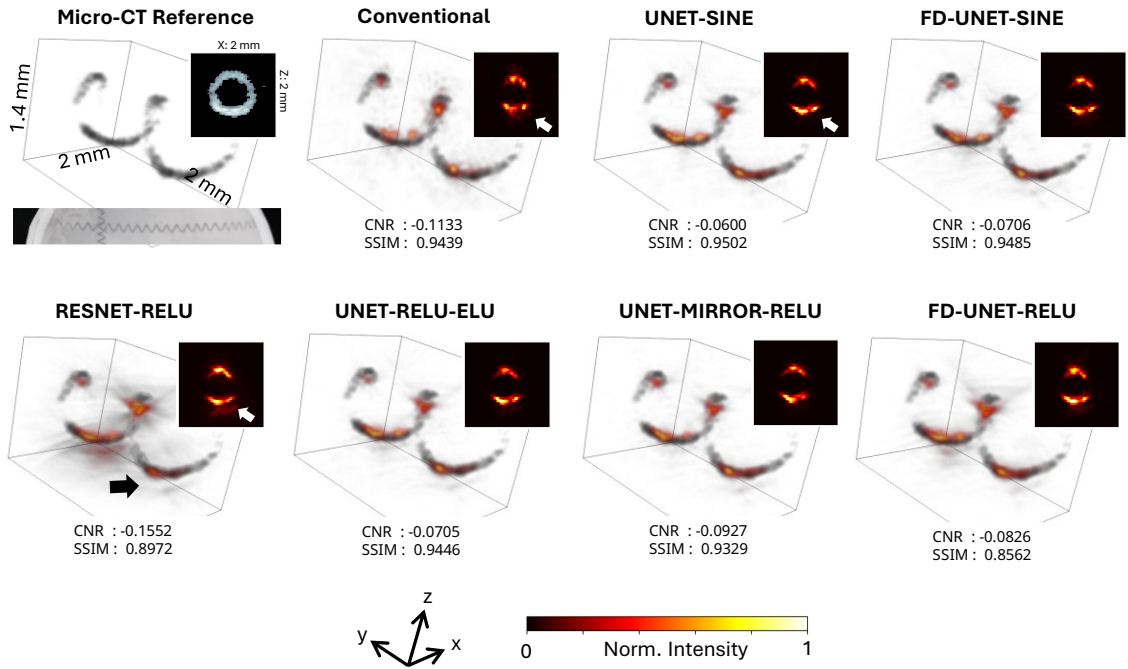

**Fig. 8.** Quantitative evaluation of 3D Spiral Phantom. Arrow pointing at the intensity outside the spiral structure. ROI on Common Signal Processing highlighted the missing vertical component.

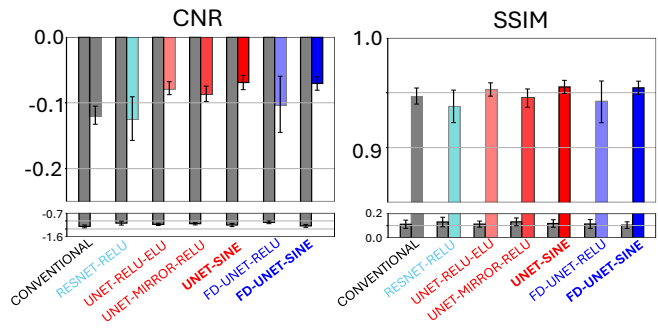

**Fig. 9.** Quantitative evaluation of leaf phantom study. Gray and colored graph are without and with CF, respectively.

Fig. 10 displays a snapshot of the microvasculature. As shown, models with ReLU/ELU, mirrored ReLU and sine activation function correctly reconstruct the micro vessel in accordance with the conventional. We highlighted the strong contrast of the bifurcating structure on the DL enhanced reconstruction, except when using all ReLU activation functions (i.e, ResNet and the FD-UNET). As investigated by the FWHM, the represented vessel also has a proper tube-like structure with uniform diameters in XY and XZ plane.

In the phantom experiment (Fig. 5), all models showed intensity leakage near bifurcations and lowering the CNR compared to the conventional. The sine-activated model was least affected, showing the highest CNR among models. However, the sine-activated model tended to suppress small details while exposing larger vessels to deeper skin as seen in Fig. 10. This trade-off matches the phantom results, where sine activation function favored larger structure.

To elucidate continuity of the enhanced vessel, we registered the volumes into a large FOV. Prior to registration, each volume was expanded to 3 mm³ (128 voxel side length) to continuously zero-pad volume larger than actual FOV without changing the resolution. Fig. 11 (next two pages) showed the comparison of the large FOV reconstructed by the UNET models. In general, all reconstruction lack of vertical microvasculature caused by the quarter-π arrangement of the sensor which was not alleviated by the present DL model. The sites in Fig. 11 highlighted a key difference among models, i.e., the sine activated model strongly exposed larger vessels. This made the FOV of the sine activated model focused on deeper regions where larger vessels were prominent. This behavior is considered as a trade-off since the removal of such small unstructured capillaries (mostly superficial) had contributed to the higher CNR of the sine activated model.

## IV. DISCUSSIONS

### A. Training via Random Sub-resolution Spherical Absorbers

We explored a method to create training datasets using point absorber model. We validated this approach in 3D using both Micro-CT and PAT imaging of the same target, from phantom to *in vivo*. Our results showed that the point absorber models are effective for training, aligned with recent work that represents the 3D PAT volume with spheres [38]. This suggests that complex datasets like real microvasculature may not be essential and allows for flexibility in training data.

Regarding the choice of training distribution, our 3D-PAT with visible wavelength was limited to the optical diffraction limit at ~ 2 mm below the skin [39]. Further, the limit of the information gathered by the hemispherical transducer is at FOV of 2 mm³ around the focal point. At this confined FOV of the superficial skin, the absorbers are mostly red blood cells flowing inside narrow impersistent capillaries [40], [41]. This was depicted on the *in vivo* study where large vessels outside the 100-um-radius training distribution rarely appear. Therefore, we had chosen a radius distribution of the spherical absorbers that also matched the actual imaging.

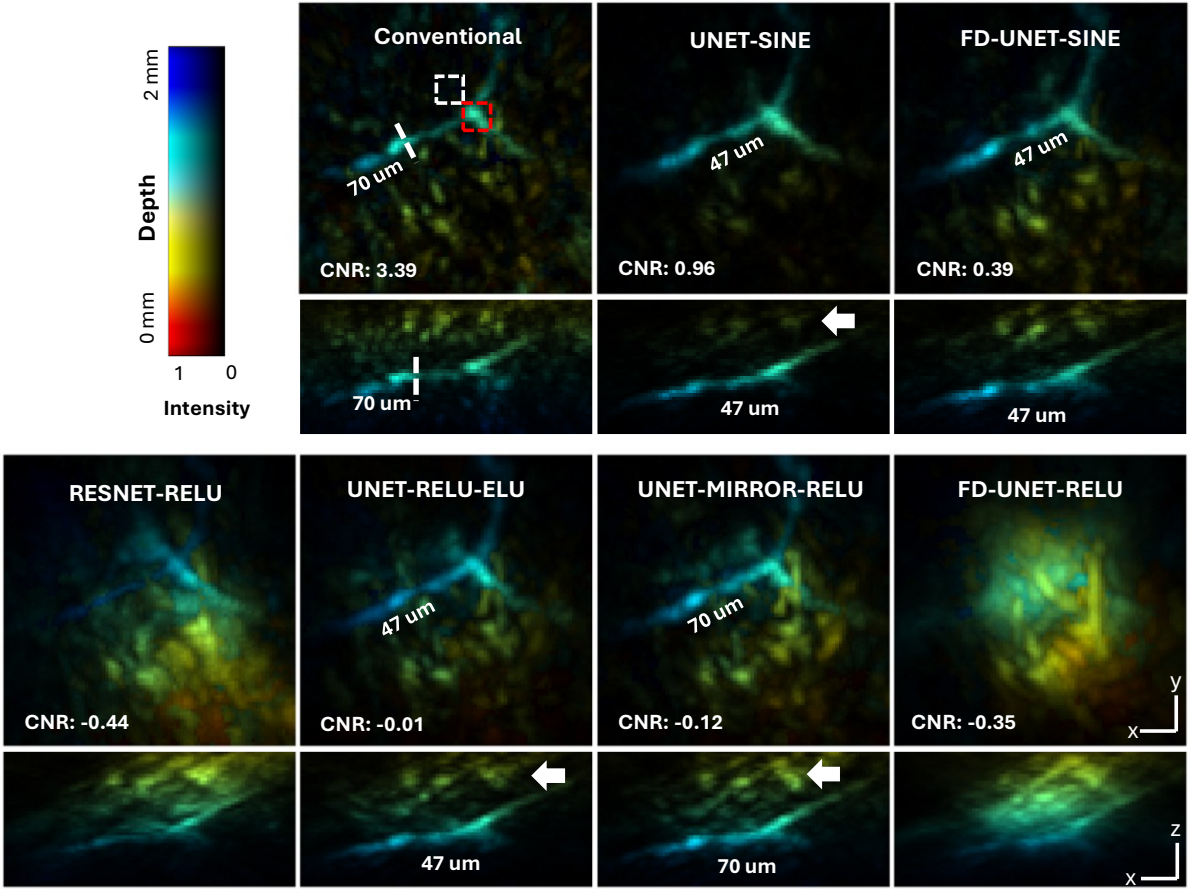

**Fig. 10.** *In vivo* evaluation of different model. Red and white ROI on Common Signal Processing represents signal and noise ROI, respectively. Arrow pointing at missing intensity at superficial skin. XY and XZ scalebar is 250 um.

Finally, we discuss that our simulation lacks the support of inhomogeneous speed of sound. Studies explore the use of DL for estimating the speed of sound [42], [43], particularly on regions where tissues are much differed and bending the wave propagation path e.g., imaging brain with skull intact [44]. We found work by [44] provided a theoretical basis for this acoustic bending which could be incorporated into our simulation. We acknowledge the limit and suggest that k-Wave is still a superior simulator in this regard. Nonetheless, we did not consider the inhomogeneous speed of sound because most of the present model [14], [15], [16] indeed assumed the same homogenous speed of sound and did not fully utilize k-Wave.

### B. Re-emergence of the importance of lower spatiotemporal spectrum on the implicit filter behavior

The experiments with Gaussian noise showed that the trained model acts as an implicit filter. We found that the noise responses were effectively expanding the spatiotemporal spectrum of the original data. This supports recent work on implicit neural representations that use oscillating functions (e.g., sine as activation [31]) to capture high-frequency details [45]. We were able to confirm such implicit behavior which might be attributed to the randomization during training.

As we discussed, all models had implicit behavior that enhanced low spatiotemporal spectrum that potentially belonged to two profound notions: the spectral bias of neural network [33] or the excellence of the model in recognizing the importance of low spectrum component of the PARF [46], [47]. It is well-known that neural networks favor learning lower spectrum due to their robustness to random perturbations. However, our training stopped at which the models had already been converged (See Supp. Mat C), meaning that the learning capacity was reached and hence the model indeed favors lower spectrum until the end of the training period. Such behavior in retaining lower spectrum then can only be explained by nature of the PARF where it is often diminished by the bandwidth limiting properties of the sensor [46], [47].

However, our DL strategy suffers from the same demerit as an implicit neural representation, i.e., re-training is necessary for different configurations of the transducer such as beyond twice interpolation. As supplement, training our UNET-SINE model took ~ 4.5 hours (1.5 hours / 100,000 iterations). This rather short training duration was attributed to the fact that our case of 2D DL on sensor-wise PARF had much fewer FLOPS at 256-time samples x (2 x 256 sensors) if compared to volume-to-volume 3D DL with FLOPS of 86³. Including the ease in obtaining simulation datasets, re-training and fast implementation has now become less burdening.

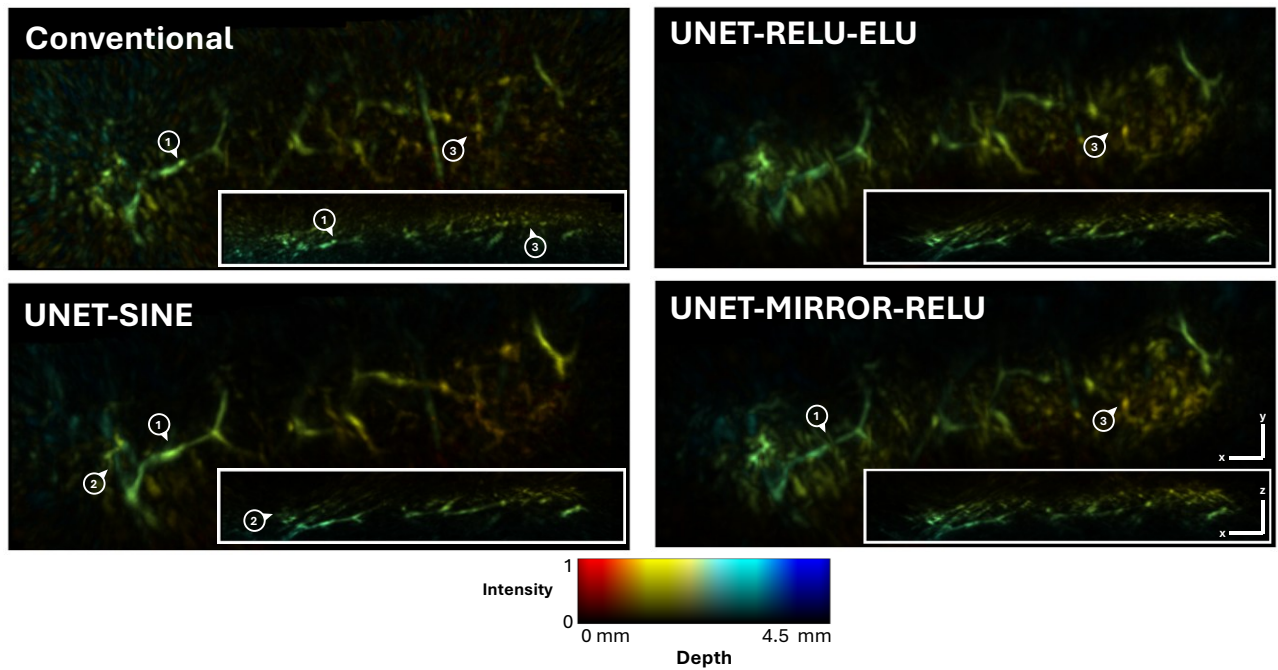

**Fig. 11.** *In vivo* large FOV shown as depth-encoded imaging in XY and XZ plane. Arrow (1) showed potential missing structure on UNET-SINE. Arrow (2) showed UNET-SINE has clean vessel visibility. Point (3) also showed similar structure on UNET-SINE and Conventional while the rest are not similar. Scalebar is 1 mm.

### C. Deep learning enhancement against conventional signal processing

We further assessed the limitations of sensor-wise PARF DL enhancement using practical results. Notably, we found that CF played a crucial role in suppressing random and out-of-axis noise [35], rather than specifically addressing streaking artefacts. In contrast, streaking artefacts typically manifest near the vicinity of the FOV and were shown to be remained after CF weighing on the example at Fig. 6. Without adequately addressing these non-streaking noise sources, the evaluation risked becoming biased and potentially obscuring the DL model's effectiveness in mitigating streaking artefacts.

Further, we discuss reduced FOV upon CF-weighing. While worsened by the CF, we identify the root cause of the FOV reduction due to using linear interpolation prior to DL enhancement as shown on Fig. 6. While stitching might help recover the reduced FOV, fully alleviating the issue would need to directly embed the interpolating function to the model.

The Wiener filter may serve as a baseline for the bandwidth issue but not necessarily for the sparsity-induced streaking artefact issue. It must be noted that the linear interpolation itself was not a standardized solution and even introduced a new problem of reduced FOV that did not exist on the conventional (see Fig. 6). Align with [48], we explored alternative methods to linear interpolation by compressed sensing with Curvelets [49]. Therefore, this compressed sensing suits well as an advanced comparator of the sparse sensor problem for our future review study.

### D. Sine activation function against others

Upon our exploration, we found that the choice of activation function was more important than the size of the model's trainable parameters. Specifically, those models powered by ReLU-only activation function were always the least performing model. In this study, all ReLU-only activation functions must be taking input range of [0, 1] (see Table I) to avoid clipping the negative intensity, hence providing fair comparison. Even though, we still found that all ReLU-activated models were underperformed. Moreover, the ReLU-activated FD-UNET did not make much improvement despite possessing larger trainable parameters than ResNet, Since the FD-UNET with sine activation function leaps the performance, we understood that the issue came from the improper choice of activation function.

As we observed across experiments, the boost in the CNR of the sine-activated model was likely caused by the less leaking intensity near the bifurcating structure. Such intensity gave cues of a known other artefact in photoacoustic (and ultrasound) named the sidelobe [50]. Unlike striking artefact that appeared around the vicinity of the FOV, sidelobe appeared near neighboring structure e.g., bifurcating structure. The observation that sidelobes strongly appeared on the non-sine-activated model and not on the common reconstruction might be related to the imperfect phase prediction.

Finally, we discuss the pitfall of the sine-activated model. A spectral defect was observed on Fig. 4 2D-FFT of the Noise Response (~wavenumber 0.2 $mm^{-1}$, frequency 23.4 MHz). We acknowledge that the current study is still lacking reason behind the source of such defects. Until the latest testing on the *in vivo* data, the defect still existed although without visible appearance on the reconstructed volume. While such missing information is not critical for visualization, processing that utilizes phase (e.g., flow photoacoustic [51]) might be significantly affected. Therefore, incorporating phase-related training objectives will be an interesting exploration in the future.

## E. Interpolation vs Extrapolation

On the theoretical exploration with Gaussian noise input, we observed that the model did not expand the spatial wavenumber, i.e., they were stuck at the same wavenumber band with the original quarter-$\pi$ acquisition. As we took equivalent structural evaluation via micro-CT, we finally observed the practical impact of such incapability in restoring the spatial wavenumber which led to missing the vertical structure of the spiral. This might also be the major reason that the micro vasculature on the *in vivo* study appeared horizontal despite the DL enhancement. Such restoration might have been fallen within extrapolation problem rather than interpolation i.e., predicting PARF signals beyond quarter-$\pi$ coverage. Studies have been conducted on this extrapolation problem [52], which might require further investigation when considering using our randomized spherical absorbing model.

## V. Conclusion

We were motivated to strive for practical usage of deep learning on 3D-PAT. We further simplified the training dataset generation by tiny, many spherical absorbers and investigated models with sine activation function. Across our extended experiments through multimodal validation and in vivo, we confirmed the feasibility, merit and demerit of our sine activated deep learning model trained from random, tiny and many spheres. Lastly, we recall that deep learning at pre-beamformed signal is significantly advantageous for 3D imaging as the data to be processed is smaller. By our *in vivo* near-real time experiment, we showed that end-to-end processing including prediction, beamforming and CF weighting, is achieved.

### Acknowledgements

A part of this study was supported by the research equipment sharing system at Tohoku University. We would like to thank Ms. Mayuko Chisiki (Graduate School of Dentistry, Tohoku University) for her technical support on the Micro-CT imaging.

## SUPPLEMENTARY MATERIALS

### A. Prefiltering Laser Artefact

Fig. A1 displayed the steps in performing pre-filtering to remove the laser artefact. Since artefacts occurred coherently at each ring, we removed the $0^{th}$ wavenumber of each ring individually. The arrow on the sensor-wise PARF pointed to the presence of the artefact on the sensor-wise PARF. Even though the intensity of the artefact on the sensor-wise PARF was low, their coherent appearance brings bright blob structure at the center axis of the volume as pointed on the arrow on the beamformed maximum intensity projection image. Since our system did not have a sensor at the center axis (i.e., hole for laser), such vertical intensity should be minimum, and thus, the trade-off between losing structures and laser interference removal is also minimum.

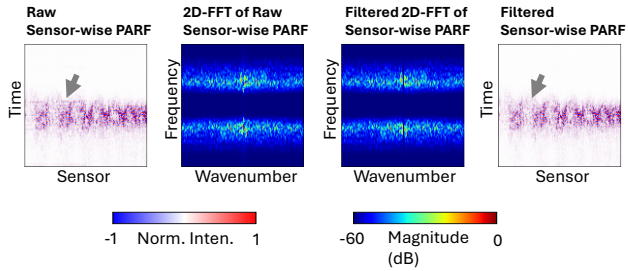

**Fig. A1.** Steps to pre-filter laser artefact.

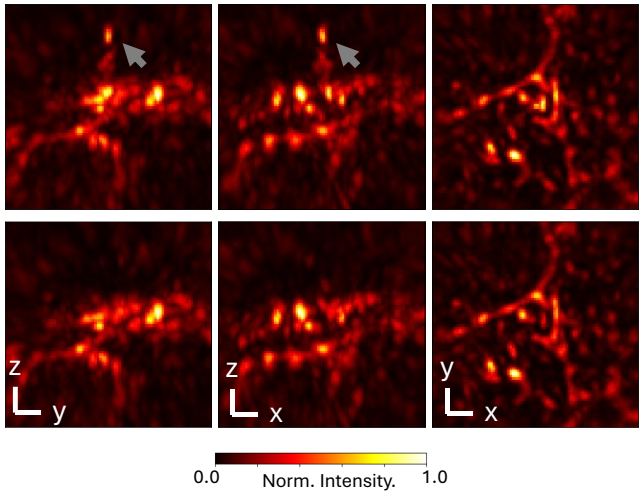

Fig. A2. Comparison before (upper) and after (lower) laser artefact removal.

### C. Preliminary evaluation

Fig. C1 displays validation by MSE across different models while training. Despite using unconventional training procedures by creating new datasets instantaneously for every iteration, we confirmed that all models were trained properly by getting converged at 100,000 iterations and the remaining 200,000 iterations were essentially fine-tuning. From the graph, we confirmed that the periodicity of the sine activation function was not impacting on the training's stability. We further explore the distribution of the pre-activated feature by inputting normalized Gaussian noise into the model. Shown on Fig. C1 (B) and (C), we clarified that all models equipped with sine activation function might be utilizing the periodicity by passing in features with distribution out of the range $[-\frac{1}{2}\pi, \frac{1}{2}\pi]$ to the sine activation function. Thus, we concluded that the training was successful by having a converging MSE and the sine activated models were correctly utilizing the periodicity possessed exclusively by the sine function.

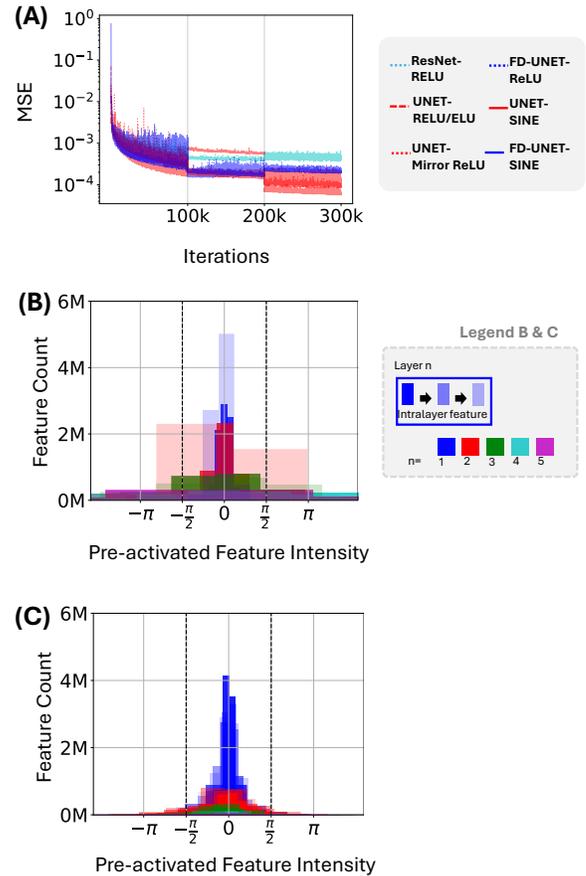

Fig. C1. Training results. (A) Training trajectory of all models. (B) and (C) are pre-activated feature distribution of UNET and FD-UNET, respectively.

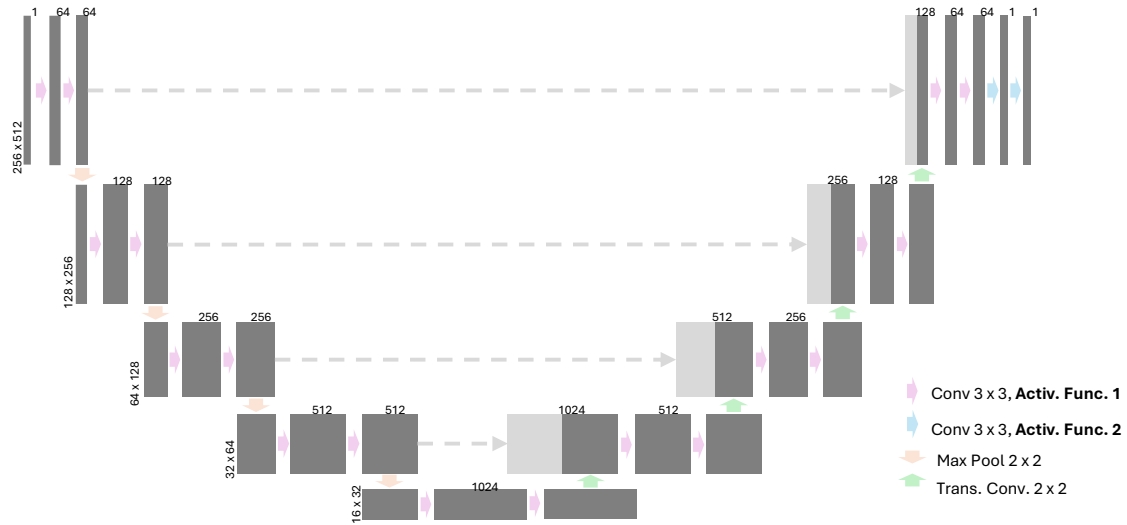

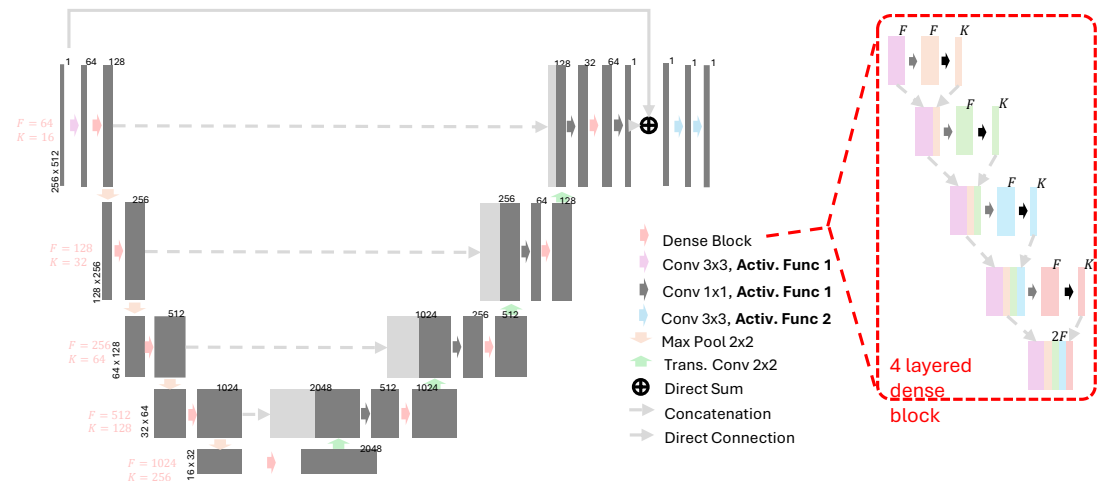

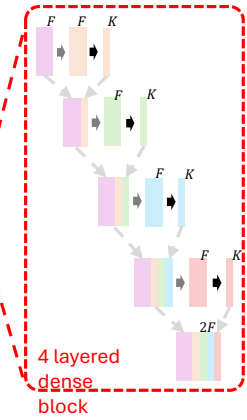

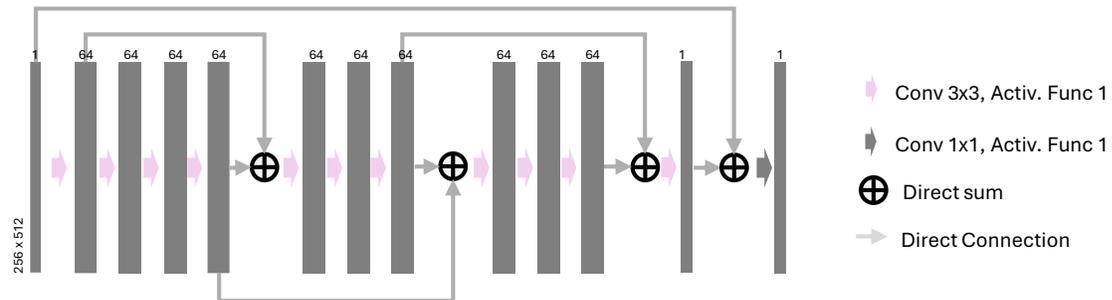

Appendix B. From top to bottom: UNET, FD-UNET and ResNet.